\documentclass{aa}

 \usepackage[dvipdf]{graphicx}
 \usepackage{epsfig}
 \usepackage{color}
 \usepackage{amsmath}
 \usepackage{lscape}  
 \usepackage{natbib}
 \usepackage{multirow}
 \bibpunct{(}{)}{;}{a}{}{,} 

\begin{document}  
  
 \title{No evidence of mass segregation in massive young clusters}
 
\author{J. Ascenso\inst{1,2}, J. Alves\inst{3} \and M. T. V. T. Lago\inst{1,4}}

\institute{Centro de Astrof\'isica da Universidade do Porto, Rua das
  Estrelas, 4150-762 Porto, Portugal \and Harvard-Smithsonian Center
  for Astrophysics, 60 Garden Street, Cambridge, MA 02138, USA \and
  Calar Alto Observatory---Centro Astron\'omico Hispano-Alem\'an, C/
  Jes\'us Durb\'an Rem\'on 2-2, 04004 Almeria, Spain \and Departamento
  de Matem\'atica Aplicada da Faculdade de Ci\^encias, Universidade do
  Porto, Rua do Campo Alegre, 657, 4169-007 Porto, Portugal}


 
 
  \abstract
  {}
  {We investigate the validity of the mass segregation indicators
    commonly used in analysing young stellar clusters.}
  {We simulate observations by constructing synthetic seeing-limited
    images of a 1000 massive clusters (10$^4$ M$_\odot$) with a
    standard IMF and a King-density distribution function.}
  {We find that commonly used indicators are highly sensitive to
    sample incompleteness in observational data and that radial
    completeness determinations do not provide satisfactory
    corrections, rendering the studies of radial properties highly
    uncertain. On the other hand, we find that, under certain
    conditions, the global completeness can be estimated accurately,
    allowing for the correction of the global luminosity and mass
    functions of the cluster.}
  {We argue that there is currently no observational evidence of mass
    segregation in young compact clusters since there is no robust way
    to differentiate between true mass segregation and sample
    incompleteness effects. Caution should then be exercised when
    interpreting results from observations as evidence of mass
    segregation.} %
 

 \maketitle
 
 
\section{Introduction} \label{sec:introduction}

The issue of mass segregation in globular and open clusters has been
discussed in the literature for over 20 years. Historically, the first
indicator of mass segregation was simply that the brightest, most
massive cluster members lay closest to the cluster core, whereas the
faintest, lower-mass members fill the whole extent of the cluster area
\citep[][and references therein]{McNamara86}. The mass segregation
issue has since been complemented with specific properties that
overall quantify the differences in the spatial distribution of high-
and low-mass stars. The most commonly used for this effect are: (1)
the radial dependence of the mass function
\citep[e.g.,][]{Moitinho97,Stolte06,Schweizer04,Gouliermis04} or
luminosity function \citep[e.g.,][]{Jones91,deGrijs02}, (2) the radial
differences in the ratio of high- to low-mass stars
\citep[e.g.,][]{Hillenbrand97}, (3) the mean mass within some
characteristic radius \citep[e.g.,][]{Sagar88,Hillenbrand98}, and (4)
the mean radius of the two distributions
\citep[e.g.,][]{Sagar88,deGrijs02} or the direct comparison of the
cumulative radial density distribution for the two subsamples
\citep[e.g.,][]{McNamara86,Tadross05}.

Even though almost all young clusters present one or more of these
properties, several authors have shown that the timescale for
dynamical relaxation is typically longer than the clusters' age
\citep[e.g.,][]{Bonnell98a}, implying that the observed mass
segregation should not be dynamical in origin. Faster phenomena, such
as violent relaxation \citep{Hillenbrand98,BinneyTremaine87} and the
fact that the massive stars have shorter relaxation times \citep[][and
references therein]{Hillenbrand98} still do not seem to be sufficient
for explaining the profusion of young clusters presenting these
properties or the implicit degree of mass segregation. The alternative
is a primordial origin, in which the distribution of massive stars in
young clusters must reflect the initial conditions and the processes
involved in cluster formation \citep[e.g.,][]{Bonnell00}.

Conversely, some authors have indirectly shown that the way some
indicators are presented is not statistically accurate. For example,
\citet{MaizApellaniz05} prove that a significant bias is introduced
when building the mass function in equal $\Delta\log(M/M_\odot)$ bins,
as is done in the literature. A statistical bias is also potentially
introduced when studying the radial properties of a cluster by
dividing the cluster area in fixed-width or constant-area annuli
rather than equal-number annuli, as each annulus will contain
different numbers of stars changing the statistical significance from
one annulus to the next.

In this paper we propose to investigate the validity of a few
traditional mass segregation indicators using synthetic, $10^4$
M$_\odot$-class clusters. We describe the biases that result directly
from the binning of the data and then explore the incompleteness in
observed samples and its consequences on the mass segregation
indicators.

\section{Simulations} \label{sec:simulations}

\subsection{Synthetic clusters}
\label{sec:synthetic-clusters}

\begin{figure*} \centering
  \includegraphics[width=15cm]{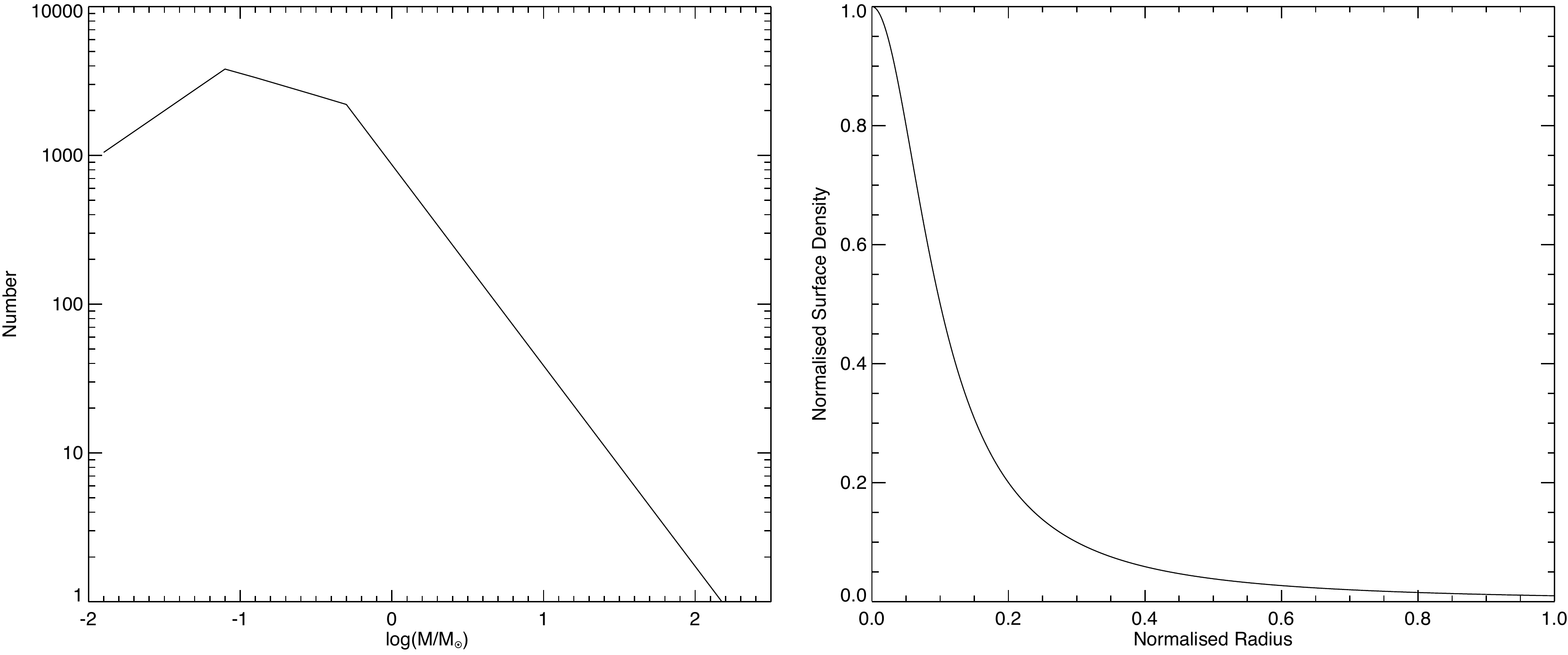}
  \caption{{\it Left}: \citet{Salpeter55} (M$ > 0.5$ M$_\odot$) and
    \citet{Kroupa01} (M$ \le 0.5$ M$_\odot$) mass function used to
    generate the synthetic clusters. {\it Right}: Normalized \citet{King62}
    surface density profile.}
  \label{fig:mf-king}
\end{figure*}

We created 1000 synthetic clusters, each containing $2 \times 10^4$
stars (total mass of $1.5\times10^4$~M$_\odot$). Each cluster member
was assigned a mass from a \citet{Salpeter55} (M $> 0.5$ M$_\odot$)
and \citet{Kroupa01} ($0.01$ M$_\odot \le $ M $ \le 0.5$ M$_\odot$)
IMF (Figure \ref{fig:mf-king}, {\it left}):

\begin{gather}\label{eq:1}
\Gamma = \frac{d\log N}{d\log M}=
\begin{cases}
-1.35 & 0.5 \leq M/M_\odot \\
-0.3 & 0.08 \leq M/M_\odot < 0.5 \\
+0.7 &  0.01 \leq M/M_\odot < 0.08.
\end{cases}
\end{gather}

\noindent The radial position (distance to the centre) of each
artificial star was drawn randomly and {\it independently of mass}
from a \citet{King62} radial surface density profile (Figure
\ref{fig:mf-king}, {\it right}):

\begin{equation}
  \label{eq:2}
  \Sigma(r)=\frac{\Sigma_0}{1+(r/r_c)^2}
\end{equation}

\noindent with a core radius $r_c$ of $0.2$ pc, using the Monte Carlo
method of the cumulative distribution function. For a cluster of $2
\times 10^4$ stars, this yields a central projected density $\Sigma_0$
around $10^4$ pc$^{-2}$. In this way the synthetic clusters are blind
to any correlation between position and mass, {\it i.e.}, they are not
mass segregated. We have not set any boundary conditions regarding the
physics of star formation or dynamical evolution, except for the
underlying assumption that the probability distribution function from
which the positions and masses are drawn are a King profile and a
Kroupa-Salpeter IMF. We have also not set any high mass cutoff.

\subsection{Synthetic observations}
\label{sec:synth-observ}

\begin{figure*}[t!]
  \centering
  \includegraphics[width=15cm]{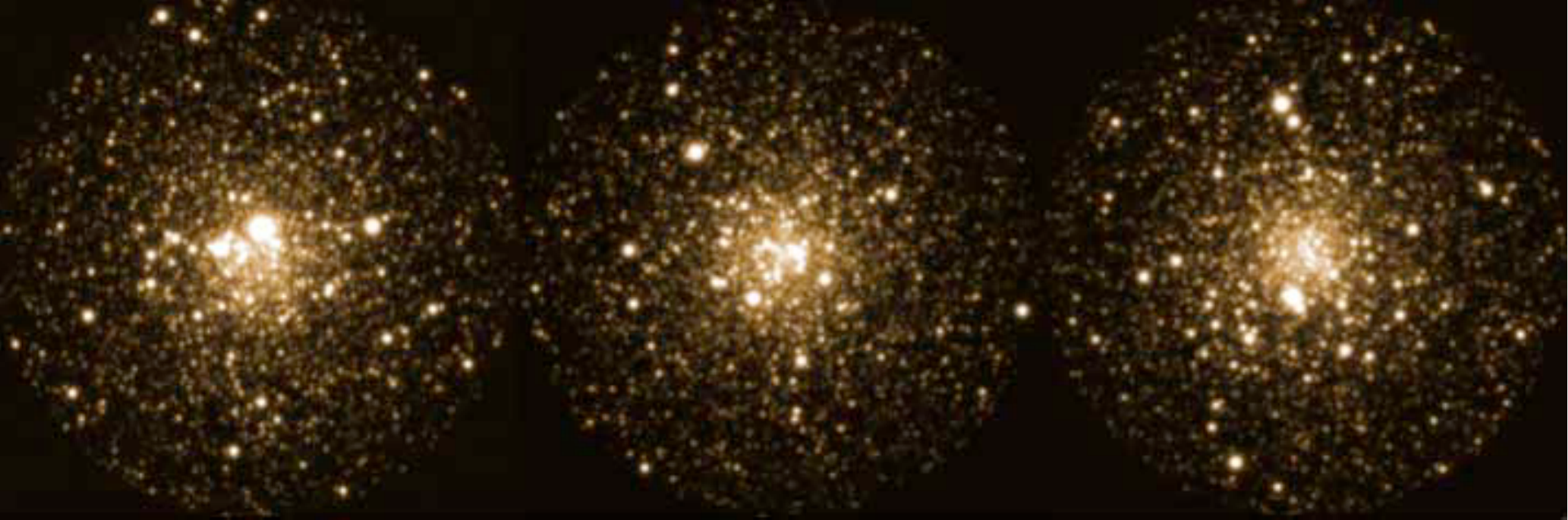}
  \caption{Seeing limited images of three simulated clusters. The
    brightness of the sources corresponds to the K-band.}
  \label{fig:images}
\end{figure*}

To investigate the impact of incompleteness due to crowding - a strong
limitation in most studies of mass segregation in real (massive)
clusters - we have used IRAF {\bf mkobject} to build ``seeing
limited'' images of the synthetic clusters. The masses were
transformed into K-band luminosities using the mass-luminosity
relation from \citet{Ascenso07} for a distance of 3 kpc and no
interstellar extinction. This configuration produced stars up to
magnitude 19. Since the Monte Carlo algorithm for the positions only
generates the $r$ polar coordinates, a value for $\theta$ was assigned
to each $r$ from a uniform random distribution between $0$ and
$360\degr$. Figure \ref{fig:images} shows three of the clusters
obtained in this way.

These images were treated as actually observed clusters, in the sense
that they were subjected to a source extraction algorithm (IRAF {\bf
  daofind}), PSF photometry (IRAF {\bf allstar}), and cuts in
photometric errors to produce the final samples. These synthetic
observations were only sensitive to stars up to magnitude 16.5, with
only $\sim$ 29\% of the original sources up to this magnitude being
detected. The 1000 catalogues produced in this way are hence
incomplete sub-samples of the original synthetic clusters. Since they
are meant to pose as real observations, we will hereafter refer to the
(incomplete) sub-samples as {\it observed*}, always maintaining the
asterisk to avoid confusion with actual observations that are not
presented here. The properties of the (complete) clusters originally
generated by the simulations will be labeled as {\it ``true''}, since
they refer to all the stars.

\section{Results} \label{sec:results}

We tested the most commonly used mass segregation indicators on
synthetic, non-segregated clusters to investigate how the way we
approach observational data may influence our perception of mass
segregation in massive clusters.

For each indicator we investigate: (1) the results expected for a
non-segregated cluster, (2) the statistical effects of binning, and
(3) the effects of incompleteness of the sample due to crowding. The
first item is measured directly from the synthetic clusters and
averaged over the whole set to produce the expected properties of a
``perfect cluster''. The second concerns the way the quantities are
represented and how it may affect the analysis. The third is measured
on the observed* clusters to explore in which ways the incompleteness
of the observed samples due to crowding can mimic the effects of mass
segregation.

We used the full width at half maximum of the stars in the simulated
observations (5 pixels) as the (arbitrary) unit of length.

\subsection{Slope {\it vs} radius} \label{sec:slope-vs-radius}

The variation of the mass function (MF) with radius is already a
traditional diagnosis tool for mass segregation
\citep{Moitinho97,deGrijs02,Stolte02,Gouliermis04,BonattoBica05,Stolte05,Bica05}. In
a mass segregated cluster we expect to find an increase in the number
of massive stars with respect to the number of low-mass stars toward
the centre, which translates into a flattening of the mass function or
an increase of the high-mass end slope, hereafter referred simply as
slope or $\Gamma$. Conversely, in a non-segregated cluster we expect
the slope to be constant with radius.

\subsubsection{Binning effects}
\label{sec:binning-effects}

This section discusses the MF slope analysis performed on the original
$2\times10^4$-star clusters.

When investigating the radial dependence of the mass function we must
bin the data twice: first we divide the cluster area into concentric
annuli and then bin the masses of the objects in each annulus to
produce the mass function for that annulus. In order to keep the
statistical significance and avoid biases, these bins should be
defined such as to keep the number of stars per bin
constant. Historically, the radial bins are defined as fixed-width or
fixed-area annuli, whereas the mass bins are defined from the
histogram of $\log(M/M_\odot)$, as constant $\Delta\log(M/M_\odot)$
bins, neither one keeping the number of stars per bin constant.
Instead, the radial bins should be defined as equal-number annuli, and
the mass function as a histogram with each bin containing the same
number of stars and divided by the resultant bin width
\citep{MaizApellaniz05}.

\begin{figure*} \centering
  \includegraphics[width=15cm]{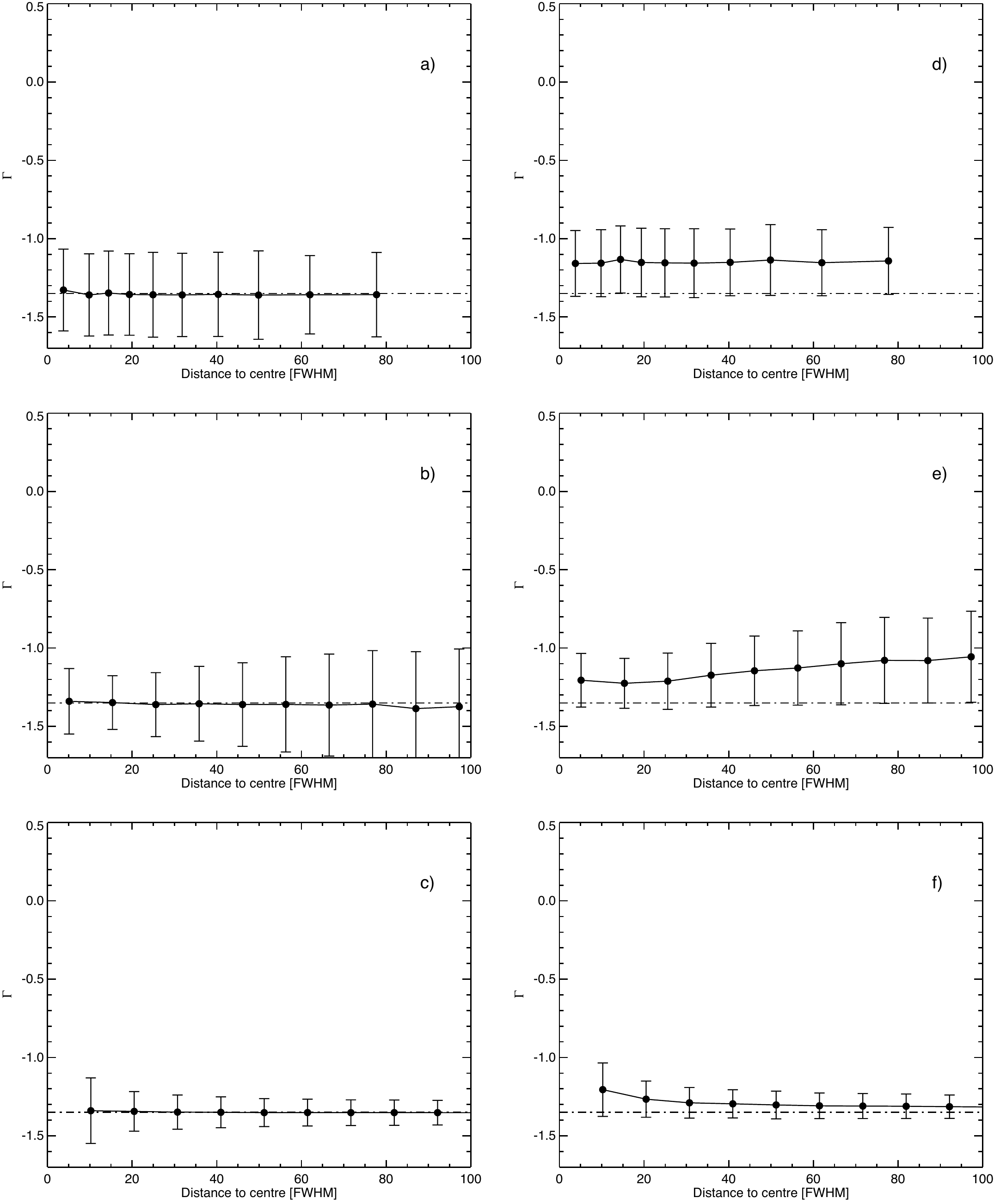}
  \caption{Mass function slope ($\Gamma$) as a function of radius for
    the synthetic clusters. {\it Panel a)}: $\Gamma$ is measured in
    equal-number (100 stars) radial annuli (the radii mark the centre
    of the annuli) and defining the mass functions such that all the
    mass bins have the same number of stars. {\it Panel b)}: $\Gamma$
    is measured in fixed-width (0.2 pc) radial annuli (the radii mark
    the centre of the annuli) and defining the mass function as in
    panel a). {\it Panel c)}: $\Gamma$ is measured in concentric
    circles (the radii mark the limits of the circles) and defining
    the mass function as in panel a).  {\it Panel d)}: $\Gamma$ is
    measured in equal-number (100 stars) radial annuli and defining
    the mass functions as fixed $\Delta\log(M/M_\odot)$
    histograms. {\it Panel e)}: $\Gamma$ is measured in fixed-width
    (0.2 pc) radial annuli and defining the mass function as in panel
    d). {\it Panel f)}: $\Gamma$ is measured in concentric circles and
    defining the mass function as in panel d).}
\label{fig:slope_panel}
\end{figure*}

Figure \ref{fig:slope_panel} shows the variation of $\Gamma$ with
radius for the 1000 synthetic 2$\times10^4$-star clusters, calculated
with several combinations of radial and mass bins. The three panels to
the left have the masses appropriately binned, according to
\citet{MaizApellaniz05}, whereas the panels to the right are produced
using the more traditional mass functions from fixed
$\Delta\log(M/M_\odot)=0.2$ histograms of $\log(M/M_\odot)$. In the
two top panels the radii are binned in annuli with equal number (100)
of stars, the middle panels have fixed-width (0.2 pc) radial annuli,
and in the bottom panels the slope of the MF is measured
(cumulatively) in circles.

The profile in {\it panel a)} is unbiased since we guarantee the same
statistical significance in both the radial and the mass bins by
keeping the number of stars in each bin constant. As such, we find
that $\Gamma$ is constant with radius, as expected for non-segregated
clusters, and equal to the input value for the simulations,
-1.35. When we change the statistical significance of the radial bins
by considering fixed-width annuli ({\it panel b)}) or circles ({\it
  panel c)}) while keeping the statistical significance of the mass
bins, we still find the behaviour expected of non-segregated clusters.

Conversely, all panels to the right display odd trends not reflecting
the conditions set for the simulations. {\it Panel d)} presents a
$\Gamma$ that is constant with radius, hence not suggesting mass
segregation, but larger than the input value of -1.35, illustrating
the intrinsic bias in characterising a MF built from fixed
$\Delta\log(M/M_\odot)$ bins \citep{MaizApellaniz05}. In {\it panels
  e)} and {\it f)} this bias conspires to produce contradictory
behaviours: {\it panel e)} shows a flattening of the MF outward,
whereas {\it panel f)} shows a flattening of the MF inward. The
profile in {\it panel f)} is what we would expect to find in a
mass-segregated cluster, although it only appears as a consequence of
the mass binning. Furthermore, as can be seen from the last bin, the
overall mass function of the cluster as measured in fixed
$\Delta\log(M/M_\odot)$ bins comes out shallower than Salpeter,
revealing a fundamental underlying bias in this representation of the
mass function, as we built the clusters to be Salpeter in the first
place.

This shows that the mass function slope is robust against radial
binning, only if the mass function itself is built in an unbiased way,
namely using the \citet{MaizApellaniz05} method to bin the masses.

\subsubsection{Incompleteness effects}
\label{sec:incompl-effects}

\begin{figure*} \centering
  \includegraphics[width=15cm]{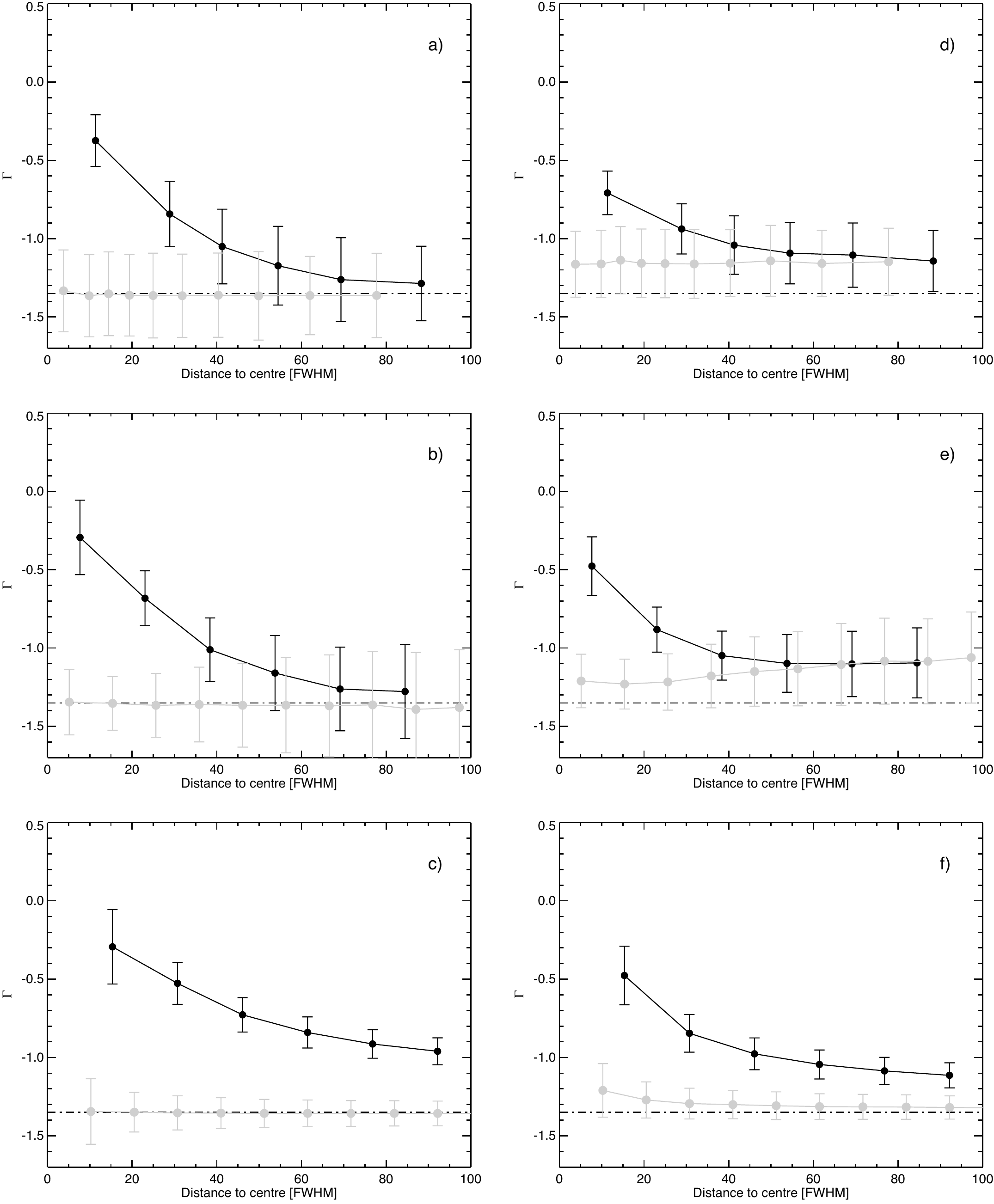}
  \caption{Mass function slope ($\Gamma$) as a function of radius for
    the observed* clusters ({\it dark symbols}) when compared to the
    ``true'' clusters ({\it light symbols}). {\it Panel a)}: $\Gamma$
    is measured in equal-number (100 stars) radial annuli (the radii
    mark the centre of the annuli) and defining the mass functions
    such that all the mass bins have the same number of stars. {\it
      Panel b)}: $\Gamma$ is measured in fixed-width (0.2 pc) radial
    annuli (the radii mark the centre of the annuli) and defining the
    mass function as in panel a). {\it Panel c)}: $\Gamma$ is measured
    in concentric circles (the radii mark the limits of the circles)
    and defining the mass function as in panel a).  {\it Panel d)}:
    $\Gamma$ is measured in equal-number (100 stars) radial annuli and
    defining the mass functions as fixed $\Delta\log(M/M_\odot)=0.2$
    histograms. {\it Panel e)}: $\Gamma$ is measured in fixed-width
    (0.2 pc) radial annuli and defining the mass function as in panel
    d). {\it Panel f)}: $\Gamma$ is measured in concentric circles and
    defining the mass function as in panel d). The profiles for the
    incomplete, observed* clusters mimic the effects of mass
    segregation.}
\label{fig:slope_panel_obs} \end{figure*}

The effects of incompleteness on the radial distribution of the mass
function slope were tested on the observed* clusters. The completeness
assessment and tentative corrections will be addressed in
Sect. \ref{sec:compl_analysis}. Figure \ref{fig:slope_panel_obs} shows
the radial dependence of $\Gamma$ for these clusters using the same
binning combinations as described above. The {\it light lines}
correspond to the radial profiles for the ``true'' clusters from
Figure \ref{fig:slope_panel}.

In all panels, regardless of the binning in radius or mass, we find a
flattening of the MF toward the centre of the cluster, a signature
typically attributed to mass segregation. These profiles are in all
similar to those described in the literature as indicative of mass
segregation
\citep[e.g.,][]{Brandl96,Stolte02,Stolte06,Schweizer04,Gouliermis04,BonattoBica05}. In
our case, since we know the ``true'' clusters are not segregated, the
finding of this signature in the observed* clusters cannot be regarded
as evidence of mass segregation in the underlying cluster, but rather
as a consequence of crowding. In the presence of incompleteness, the
statistical biases arising from binning are largely overcome by the
fact that the low-mass stars go undetected in the cluster core. The
mass-binning effects are only observed as a flatter mass function in
general.

\subsection{Ratio of high- to low-mass stars} \label{sec:ratio-high-low}

\begin{figure*}
  \centering
  \includegraphics[width=18cm]{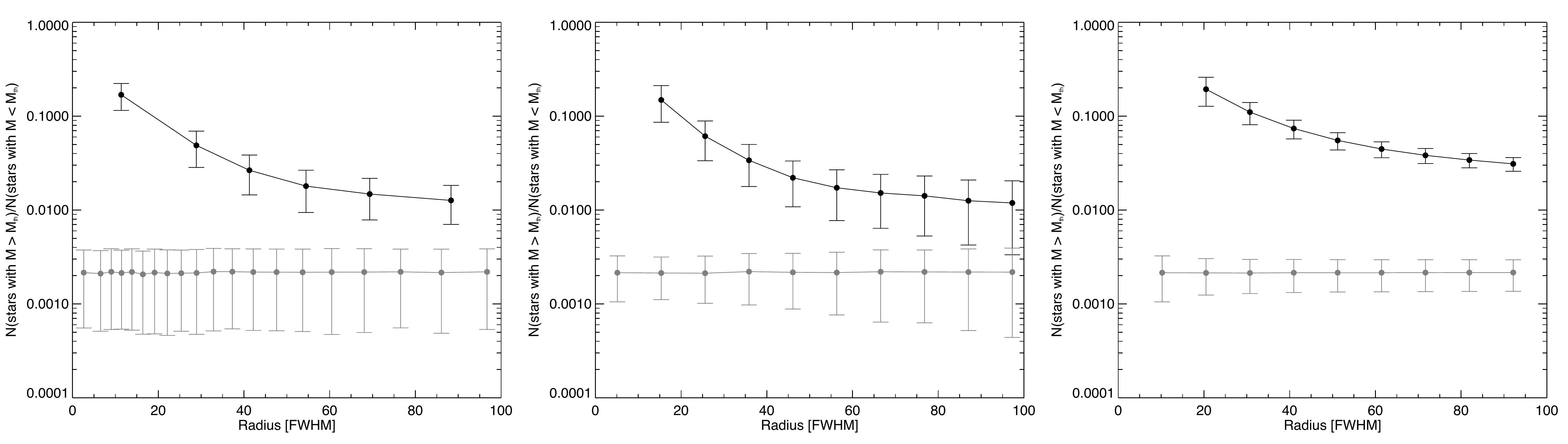}
  \caption{Ratio of high- to low-mass stars with radius for a mass
    threshold of 10 M$_\odot$ for the ``true'' clusters ({\it light
      symbols}) and observed* clusters ({\it dark symbols}) measured
    in annuli of fixed number of stars ({\it left}), fixed-width
    annuli ({\it middle}) and concentric circles ({\it right}). The
    profiles for the incomplete, observed* clusters mimic the effects
    of mass segregation.}
  \label{ratio}
\end{figure*}

In any given region of a non-segregated cluster, apart from
fluctuations, there should be the same proportion of high and low-mass
objects as imposed by the underlying mass function. In particular, the
ratio of high- to low-mass stars should not be dependent on
radius. This is indeed what we find for the synthetic clusters,
regardless of how we divide the cluster radially. The {\it light
  symbols} in Figure \ref{ratio} show this profile for a
high-mass/low-mass threshold of 10~M$_\odot$, and radial binning
consisting of equal-number annuli ({\it left-hand panel}), fixed-width
annuli ({\it middle panel}), and concentric circles ({\it right-hand
  panel}). All the profiles are flat, again validating the absence of
mass segregation in the synthetic clusters, and present no signature
of statistical biases arising from radial binning effects.

The {\it dark symbols} in the panels show the variation of the ratio
of high- to low-mass stars in the observed* clusters. For all
geometries the ratio increases toward the cluster core, suggesting an
apparent depletion of low-mass stars in the core. This is a direct
consequence of crowding that does not allow for the effective
detection of faint sources, rather than an actual absence of low-mass
stars in the underlying cluster that could be imputed to mass
segregation. Again, this profile is similar to those cited in the
literature as evidence of mass segregation
\citep{Hillenbrand97,Stolte06}.

\subsection{Mean mass} \label{sec:mean-mass}

\begin{figure*}
  \centering
  \includegraphics[width=12cm]{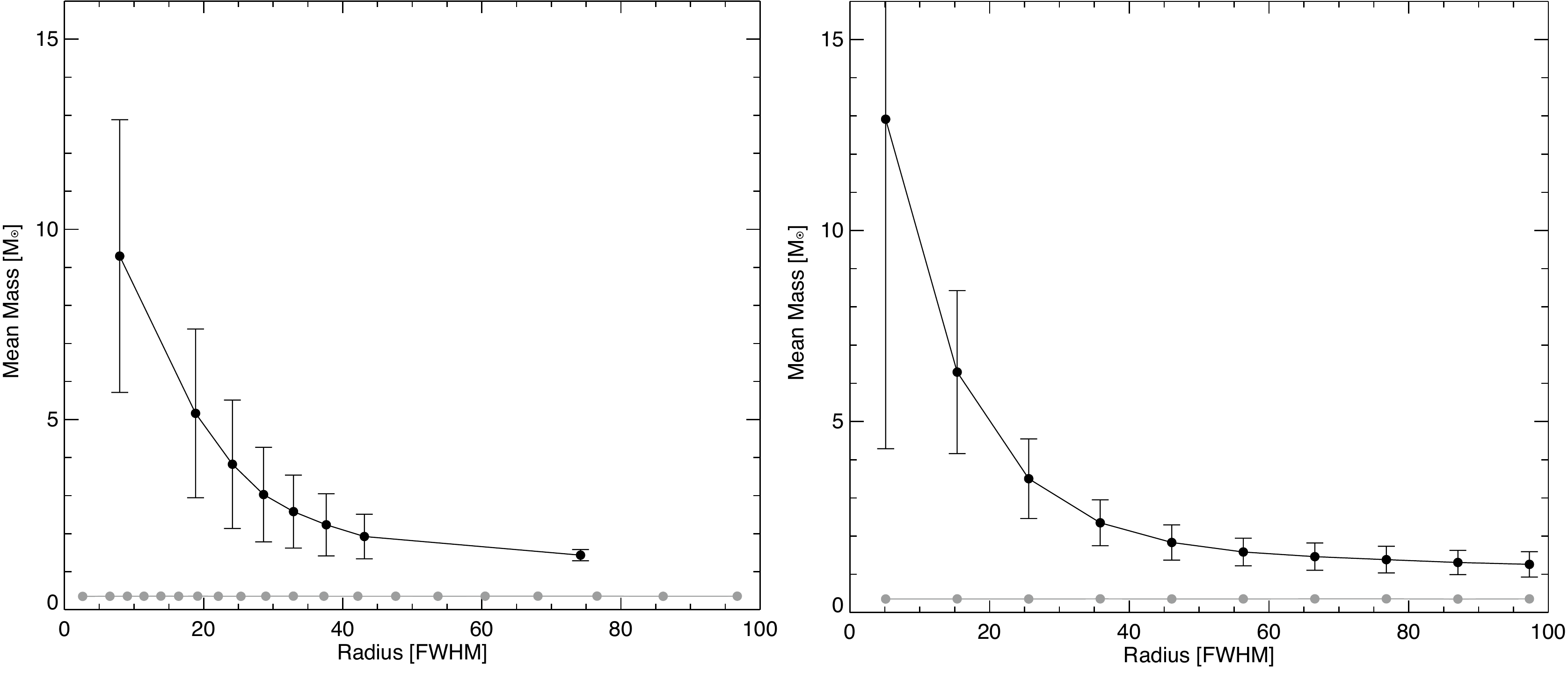}
  \caption{Mean mass of the stars within annuli of equal number of
    stars ({\it left}) and fixed-width annuli ({\it right}) for the
    ``true'' ({\it light symbols}) and observed* ({\it dark symbols})
    clusters.}
  \label{fig:mean-mass}
\end{figure*}

Following the same reasoning as before, the mean mass of a
non-segregated cluster should be independent of the region where we
choose to measure it. This is what we find when we plot the mean mass
in concentric annuli for the synthetic clusters (Figure
\ref{fig:mean-mass}, {\it light symbols}), regardless of using
fixed-number ({\it left-hand panel}) or fixed-width ({\it right-hand
  panel}) rings.

Conversely, the observed* clusters ({\it dark symbols}) display a
significant increase of the mean mass toward the cluster centre, as
the faint stars in the centre are not as effectively detected as the
massive stars, shifting the mean mass to higher values, a signature
also often attributed to mass segregation.

\subsection{Mean radius of the massive stars} \label{sec:mean-radius}

\begin{figure*}
  \centering
  \includegraphics[width=12cm]{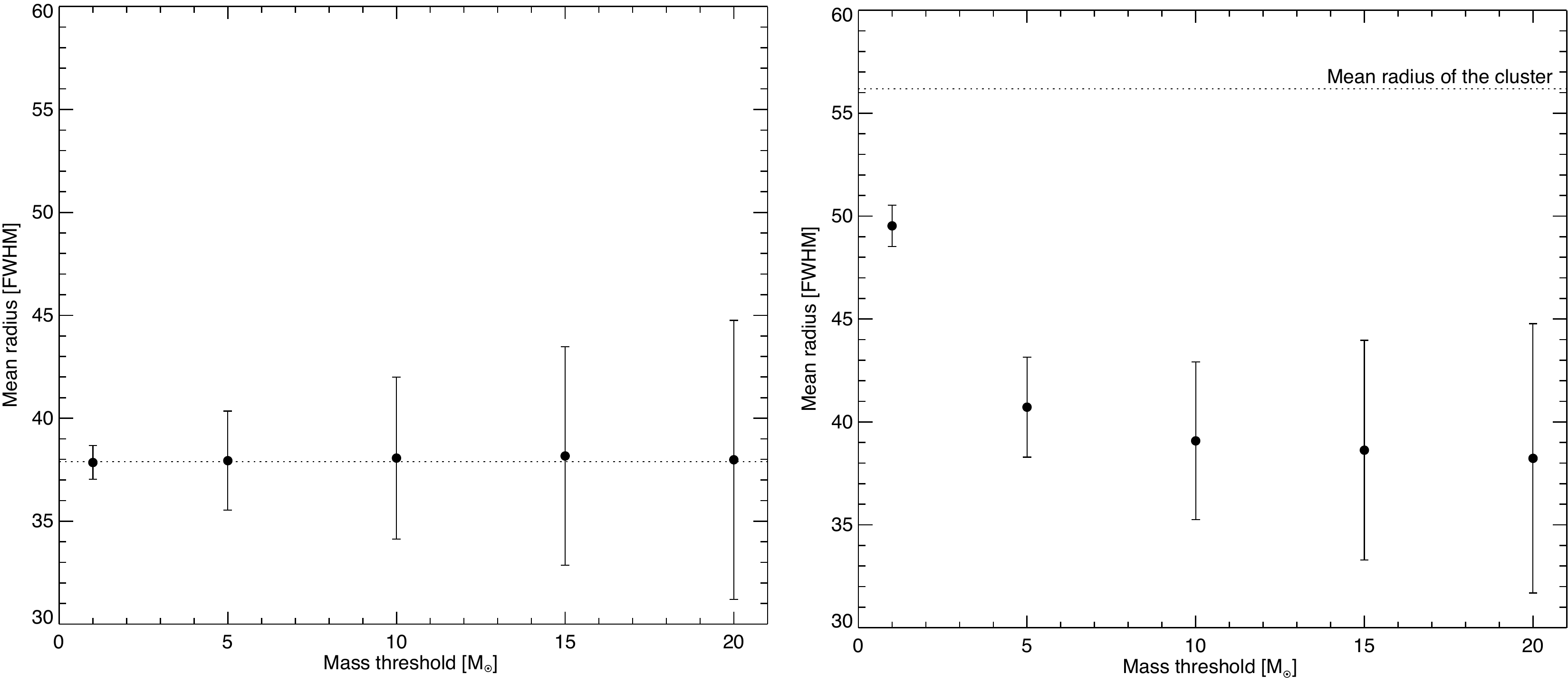}
  \caption{Mean radius of the stars with masses higher than the
    designated mass threshold for the average of the 1000 ``true''
    clusters ({\it left}) and for the observed* clusters ({\it
      right}). The {\it dotted} line represents the mean radius of the
    clusters.}
  \label{fig:mean-radius}
\end{figure*}

In the present context we define the mean radius of any sample of
stars as the mean distance of those stars to the centre of the
cluster. For each cluster, we measured the mean radius of the massive
stars and compared it to that of the cluster as a whole. The massive
star subsample was defined using mass thresholds of 1, 5, 10, 15, and
20 M$_\odot$.

We find that both the mean radius of the ``true'' clusters and that of
their massive stars have the same value for all mass thresholds
(Figure \ref{fig:mean-radius}, {\it left}), although the
cluster-to-cluster fluctuations increase with increasing threshold.
This changes for the observed* clusters ({\it right-hand panel}), as
the number of low-mass stars in the centre is significantly smaller
than before due to crowding, causing the total mean radius of the
cluster to become larger ($1.1$ pc), whereas the mean radius of the
set of massive stars remains roughly the same for almost all mass
thresholds, as these are not affected.

These profiles match those found in the literature
\citep[e.g.,][]{Sagar88,Bonnell98a,Schweizer04}, where the authors
consistently find the massive stars to have smaller mean radii when
compared to the mean cluster radii.

\section{On completeness corrections}
\label{sec:compl_analysis}

In the previous sections we have shown that incompleteness due to
crowding will mimic the effects of mass segregation in the commonly
used indicators. This is so because they have the same effect: a
depletion of low mass stars in the cluster core. The fundamental
difference is that, whereas mass segregation in young clusters implies
a physical process over which the stars of different masses are formed
or somehow appear spatially segregated, crowding simply causes the
observer to miss the low-mass stars due to the resolution limitation
of the instrumental set-up. Many authors are aware of these
limitations and apply more or less sophisticated completeness
corrections to their samples, but how good are these corrections?
Since we know in advance the exact composition of our synthetic
clusters, we used one of them to perform a thorough investigation on
the completeness assessment and correction process.

\subsection{Completeness tests}
\label{sec:compl-tests}

\begin{figure} 
  \resizebox{\hsize}{!}{\includegraphics{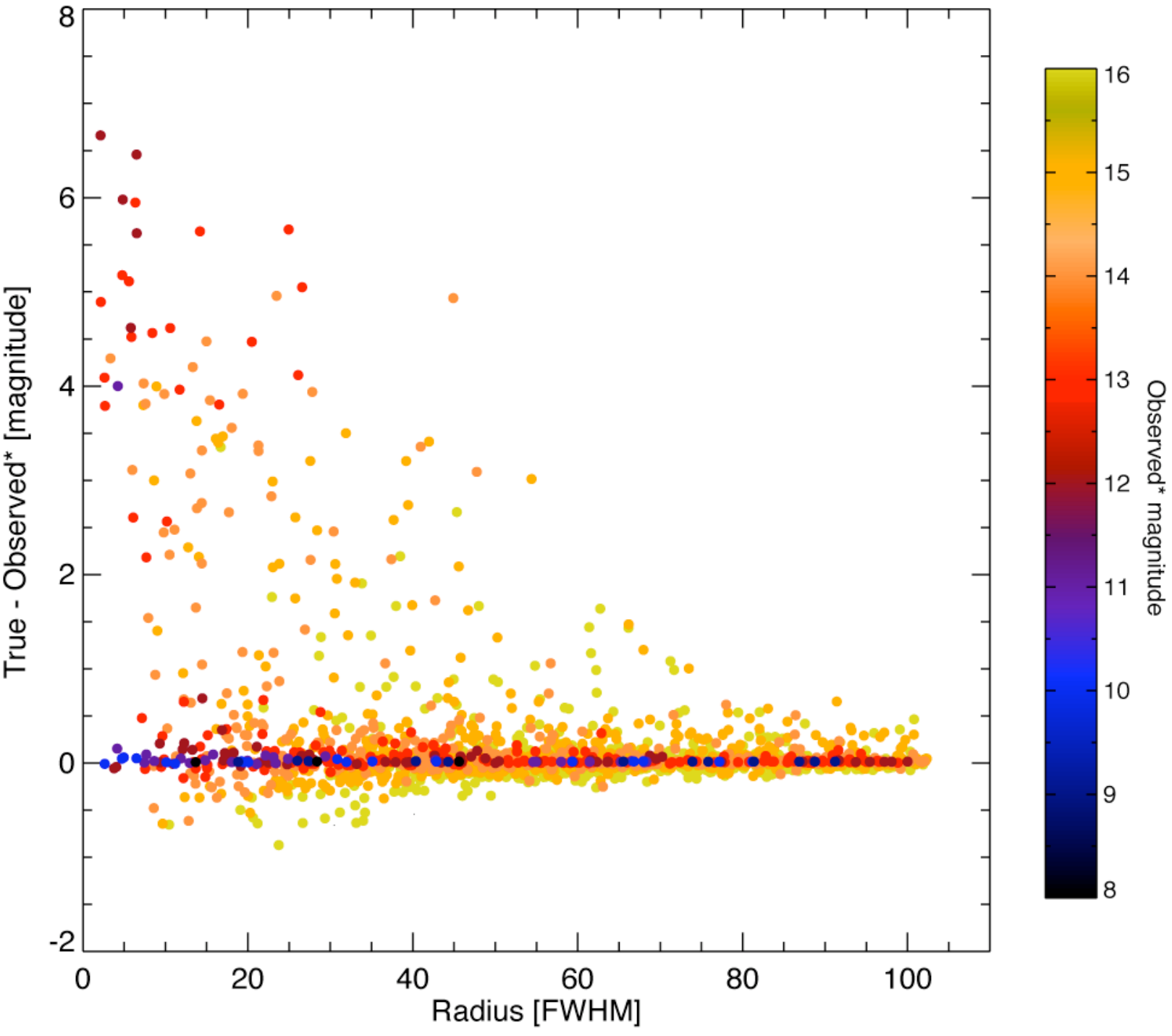}}
  \caption{Difference between the true and observed* brightness for
    the cluster stars as a function of distance to the centre. The
    colour-code maps the observed* brightness of the stars.}
  \label{fig:deltamag-radius}
\end{figure}

The direct comparison of the true and observed* properties of a
synthetic cluster is the most immediate way to gain insight into what
is actually lost to observational limitations. Figure
\ref{fig:deltamag-radius} shows the difference between the true and
observed* brightness of cluster stars as a function of distance to the
centre, while the colour-code maps the observed* brightness of the
stars. It becomes clear that the two relevant consequences of crowding
toward the cluster core are: (1) hampering source detection due to
confusion caused by the close proximity of the sources, and (2)
inflate the stars' brightness by blending their flux with that of
unresolved neighbours. As a result, as we move into the centre of the
cluster, we are less and less sensitive to the faint stars, and will
tend to overestimate, sometimes by several magnitudes, the brightness
of those we do detect. The bright stars, on the other hand, are
equally detected everywhere throughout the cluster and their measured
brightness is hardly affected by the crowding.

However insightful as this comparison may be, in real clusters we must
rely on completeness tests to determine the extent to which we may
trust the observations, as we lack the privileged information about
the cluster's true composition.  To address potential accuracy and
reliability issues of completeness tests, we computed them for a
synthetic cluster as if it were an actually observed image.  For every
0.5 magnitude bin, we added artificial stars to the image in a grid
such that each star is separated from its closest neighbour by two
times the radius of the PSF $+1$ pixel. By forcing the artificial
stars to be in such a grid we sampled the full extent of the cluster
area without adding to the crowding. The images for each magnitude
were then subject to source detection and photometry, and the output
lists of sources were compared with the input grids. The completeness
for magnitude $m_{in}$ is then defined as the number of detected grid
stars with measured magnitudes $m_{out}$ that satisfy the condition
$|m_{in}-m_{out}|<0.1$, divided by the total number of stars of
magnitude $m_{in}$ in the input grid. The latter condition implies
that an artificial star blended with a cluster star such that it
affects its magnitude beyond the reasonable photometric uncertainty is
rejected for completeness purposes, which happens very frequently in
the crowded core, mainly for the faint stars. The outcome of these
tests is therefore a high-fidelity completeness assessment that
contains information, not only about the detection success rate, but
also about how blending affects the incompleteness.  We use the
definition of completeness described above in the following sections.


\subsection{Global completeness}
\label{sec:global-completeness}

\begin{figure}
  \resizebox{\hsize}{!}{\includegraphics{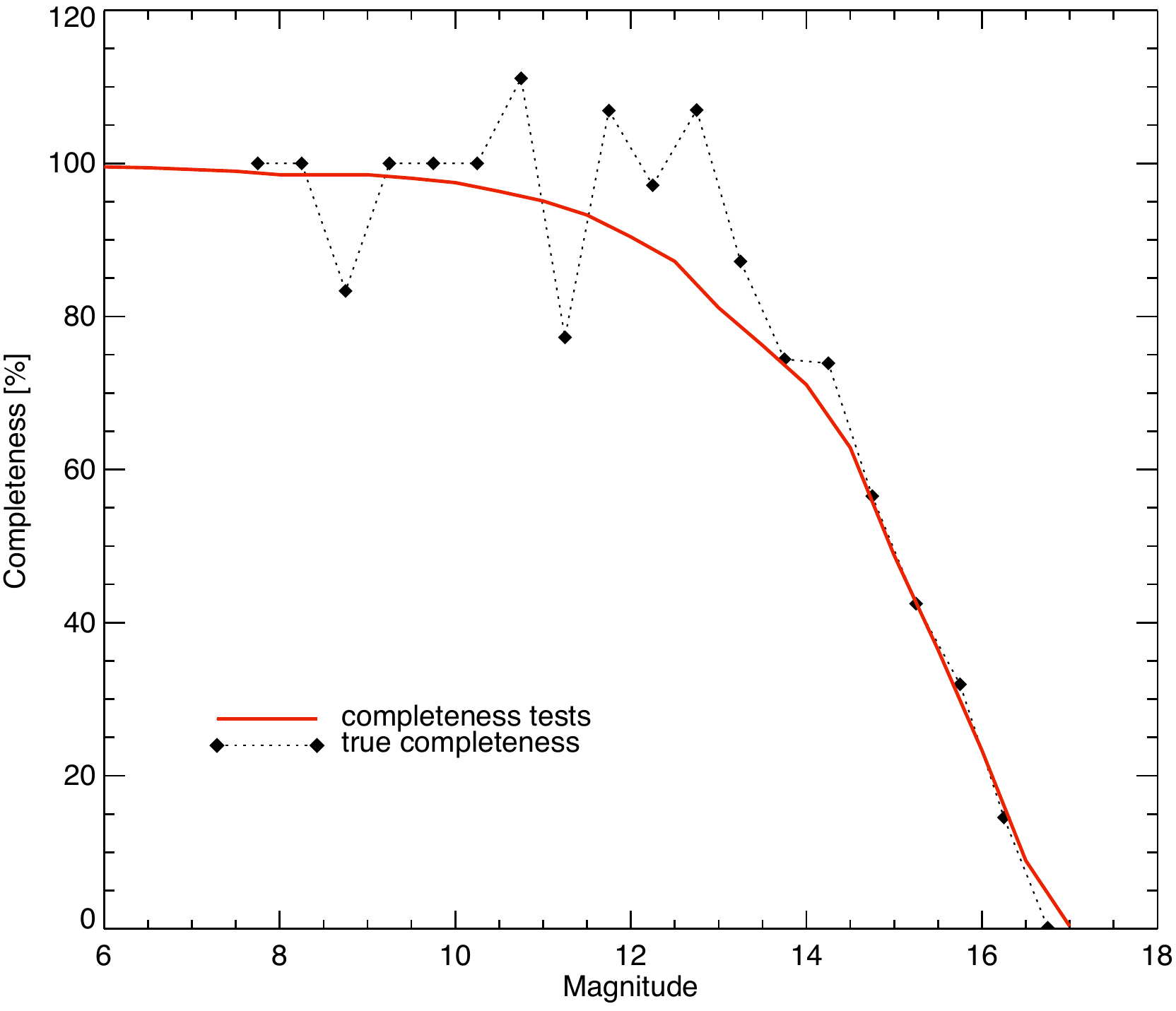}}
  \caption{Global completeness as a function of magnitude from the
    completeness tests ({\it red solid line}), and from direct
    comparison of the observed* and true brightnesses ({\it dotted
      line}).}
  \label{fig:global-compl}
\end{figure}

The red solid line in Figure \ref{fig:global-compl} shows the global
completeness - the fraction of artificial stars recovered with respect
to the input stars in the grid - as a function of magnitude. These
tests return a 90\% global completeness for magnitude 12 (6.2
M$_\odot$ in our example). The {\it dotted line} is the {\it true}
completeness defined here as the fraction of observed* cluster stars
relative to the true number of stars for each magnitude in the
synthetic cluster\footnote{A summary of the nomenclature and
  definitions used here is given in Table \ref{glossary} of Appendix
  \ref{sec:nomenclature}.}. The local disparities between the two
profiles are due to unresolved (blended) sources: whereas blending is
excluded for the purpose of completeness tests (see
Sec. \ref{sec:compl-tests}), it does occur in observations - blended
stars will appear in the list of observed* sources as single stars
with good photometry. As a consequence, blended stars ``leak'' to
different magnitude bins and cause the true completeness to be
contaminated in an unpredictable way. For this reason, and again
because the completeness tests include only single stars, the profiles
fail to match for some magnitudes. Nevertheless, the overall agreement
indicates that the accuracy of the global completeness tests is quite
reasonable.

\subsubsection{Correcting luminosity functions}
\label{sec:corr-lumin-funct}

\begin{figure}
  \resizebox{\hsize}{!}{\includegraphics{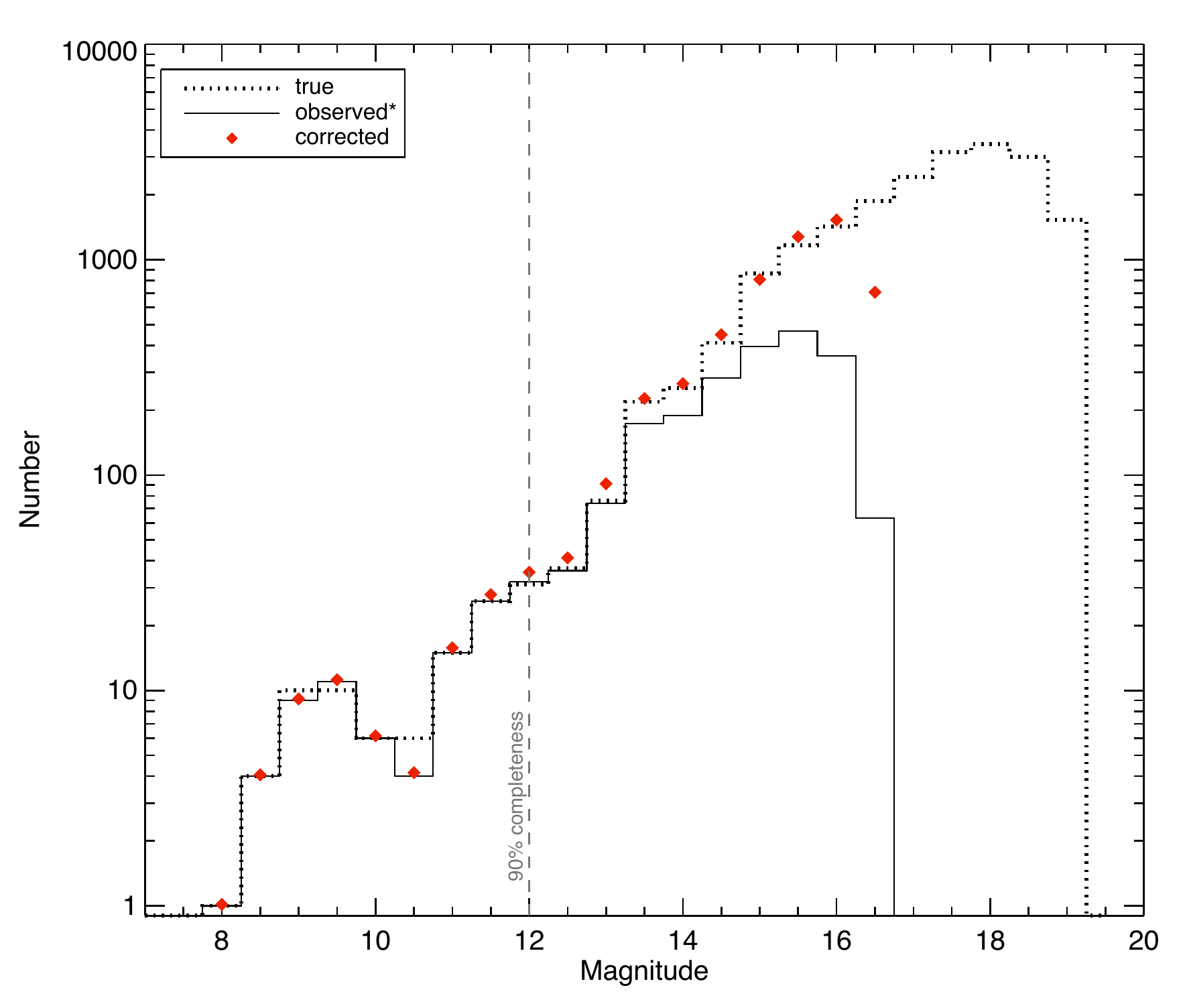}}
  \caption{Comparison of the observed* luminosity function ({\it solid
      line}) with the true ({\it dotted line}) and completeness
    corrected ({\it red diamonds}) luminosity functions.}
  \label{fig:KLFs}
\end{figure}

An important and surprising corollary of the validation of global
completeness tests above is that the global properties of the cluster,
such as its mass function, can effectively be corrected for
incompleteness due to crowding. Figure \ref{fig:KLFs} shows the
observed* ({\it solid line}), true ({\it dotted line}), and
completeness corrected ({\it red diamonds}) luminosity functions for
this cluster. The latter was derived by dividing the first by the
completeness profile ({\it red line} in Figure
\ref{fig:global-compl}), and it is indeed very faithful to the true
luminosity function for all but the last corrected bin, where the
correction drops from 23\% to 8\%. The last reliable bin is four
magnitudes fainter than the estimated 90\% completeness limit (6.2
M$_\odot$), corresponding now to a mass of 0.3 M$_\odot$ in our
example. This implies that one would, in principle, be able to see the
first break of the mass function in this cluster even though the
completeness limit is a great deal more massive.

This example shows the potential of the global completeness tests, but
we emphasise that it is only valid for clusters with the same
characteristics as the presented synthetic clusters, when crowding is
the only source of incompleteness (e.g., no extinction), and for this
particular method of evaluating completeness. Most completeness
studies in the literature, although similar, are not as thorough as
the one described here and must therefore be validated before
extending this result to other degrees of crowding and/or
observational configurations.

\subsection{Radial completeness limitations}
\label{sec:radi-incompl}

\begin{figure} 
  \resizebox{\hsize}{!}{\includegraphics{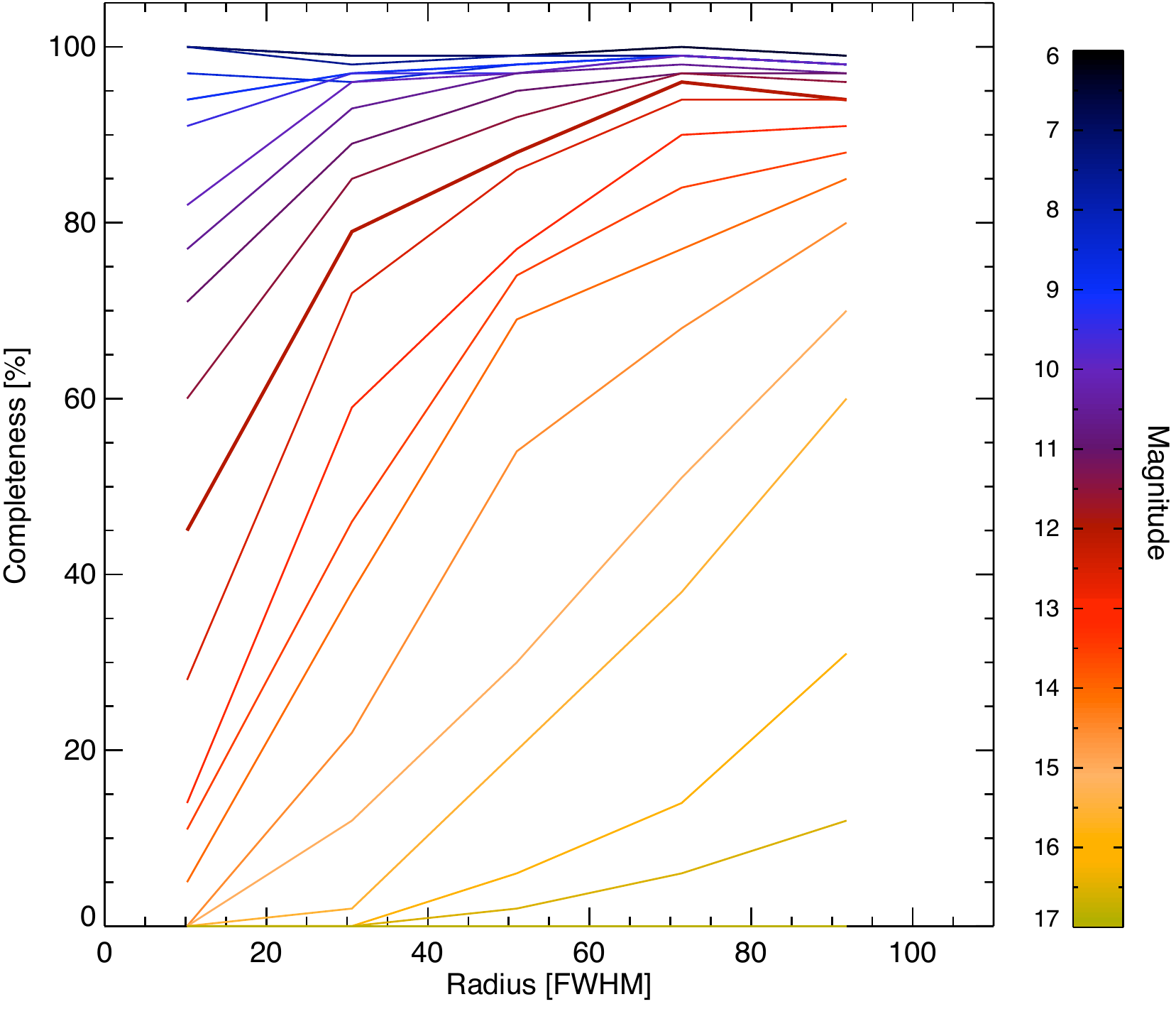}}
  \caption{Completeness tests as a function of radius for the
    different magnitudes. The line in bold indicates the 90\%
    completeness limit determined in
    Sect. \ref{sec:global-completeness}.}
  \label{fig:compl-rad}
\end{figure}

The global completeness discussed in the previous section describes
the average behaviour in the whole cluster area, but is not
representative of the cluster core where the crowding is most
severe. The completeness tests described in
Sect. \ref{sec:compl-tests} can then be analysed in concentric rings
about the centre of the cluster to estimate the radial dependence of
completeness. Figure \ref{fig:compl-rad} shows the completeness as a
function of radius for the different magnitudes.

\begin{figure} 
  \resizebox{\hsize}{!}{\includegraphics{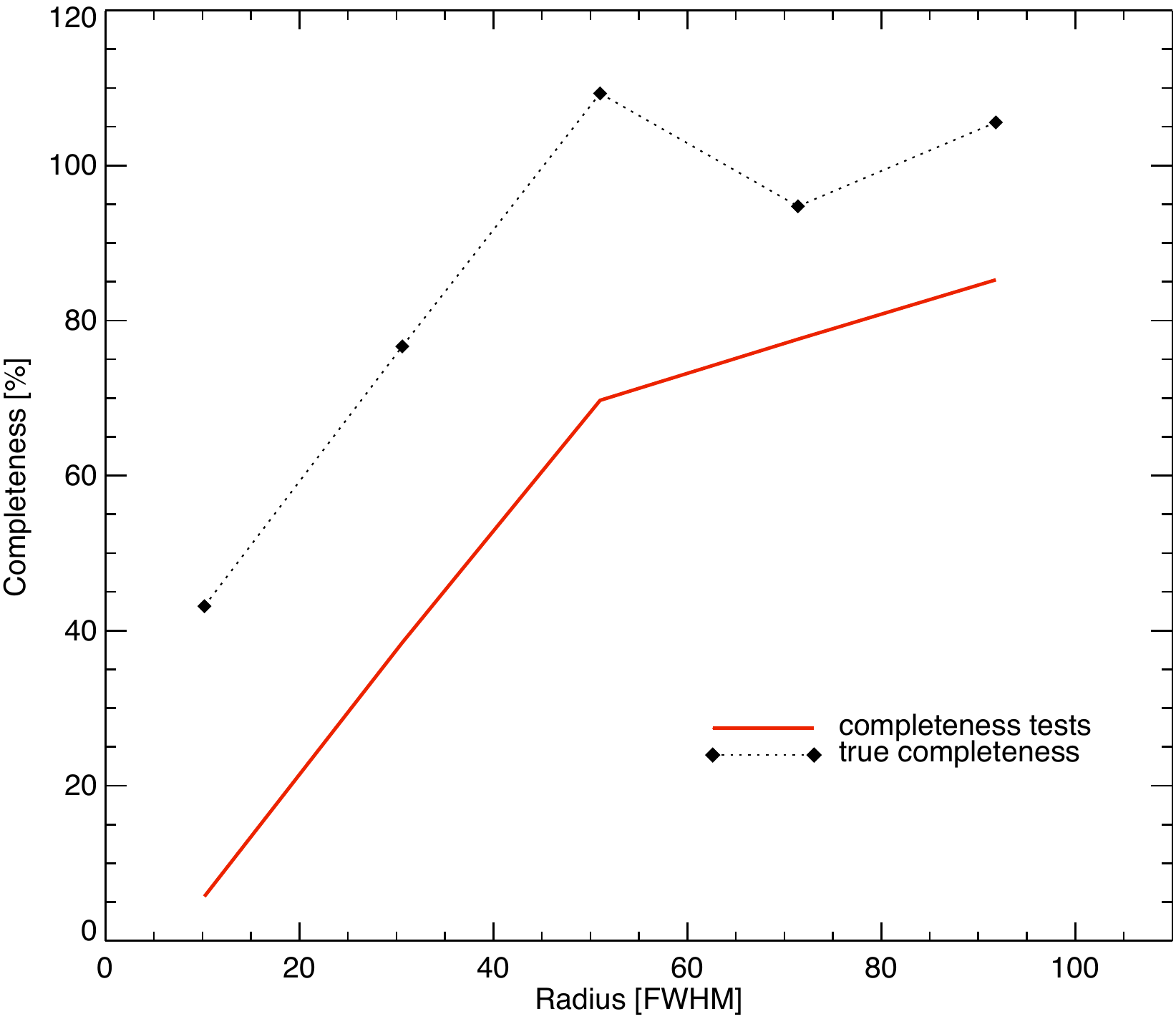}}
  \caption{Completeness as a function of radius from the completeness
    tests ({\it red solid line}), and from true completeness ({\it
      dotted line}) for magnitude 14.}
  \label{fig:true-compl-rad}
\end{figure}

Extrapolating from Figure \ref{fig:global-compl}, one would trust
these radial profiles to be a fair representation of the true radial
completeness. However, when comparing both for any given magnitude we
instead find a blunt disagreement (see Figure \ref{fig:true-compl-rad}
for magnitude 14). These differences are entirely attributable to the
blending of unresolved sources: when selecting stars of magnitude $m$
from the observed* list of sources to assess the true completeness, we
include (blended) stars that are in reality fainter, while at the same
time excluding true $m^{th}$ magnitude (blended) stars with inflated
(combined) brightnesses. The consequence is that we appear to be more
complete than what the grid completeness tests suggest simply because
we cannot differentiate between blended and single sources.

In terms of completeness corrections, these ``magnitude leaks'' due to
blending take much larger proportions than they did in the global
completeness analysis (Sect. \ref{sec:global-completeness}).  On the
global scale the effect of the magnitude leaks from the core is
largely diluted, allowing for reasonable completeness corrections.
Conversely, the radial completeness tests systematically imply very
large (over-)corrections in the cluster core, which immediately
produce a greatly inflated amount of stars, resulting in our case in
an (over-)corrected cluster with typically 3.5 times more stars (up to
magnitude 14) than the original synthetic cluster.

In terms of completeness assessment, both radial completeness
estimates in Figure \ref{fig:true-compl-rad} agree that: (1) the
global completeness overestimates the completeness in the crowded
central regions, and (2) that completeness is strongly radially
dependent being more severe in the cluster core, which ultimately
confirms the hypothesis that it is responsible for the apparent mass
segregation.

To summarise, this analysis shows that there is a radial dependence of
the completeness affecting primarily the low-mass stars that we cannot
correct for, so no radial property (such as mass segregation) can be
legitimately measured in the presence of severe crowding.

\section{Conclusions}
\label{sec:conclusions}

We used synthetic non-segregated, compact, and massive clusters to
investigate the impact of the current approach to observational data
on mass segregation studies. Our conclusion is that incompleteness due
to crowding will produce the observed properties of mass segregated
clusters, even when they are not segregated at all. Crowding causes
the massive stars to be detected more effectively than the low-mass
stars, resulting in an apparent depletion of low-mass stars in the
cluster core, which then produces the characteristics typically
attributed to mass segregation. More revealing, radial completeness
tests provide erroneous estimates of the incompleteness and, as a
consequence, lead to severe over-corrections. This is even more
unsettling if we consider that it is not possible to evaluate the
accuracy of the completeness determinations with the information from
the observations alone. This is particularly critical for distant,
rich clusters or clusters observed with poor spatial resolution or
sensitivity.

We have also found that the way to present the data may furthermore
influence the analysis, although to a much lesser extent. In what
concerns the radial study of the mass function, it is imperative that
the slope in each radial annulus be measured in mass bins with equal
number of stars, as described in detail by \citet{MaizApellaniz05}. If
this is so, then the radial binning will not influence the
analysis. The other indicators (ratio of high- to low-mass stars and
mean mass of the stars in annuli) are not affected by radial binning
effects.

This exercise showed that the study of mass segregation cannot be
dissociated from an exhaustive and rigorous study of completeness --
which is not often found in the literature -- and even then extreme
caution must be exercised when interpreting radial properties as
evidence of mass segregation.

The presence of interstellar extinction, not included in this
analysis, will affect the mass segregation indicators in a more
unpredictable way. On the one hand, the spatial distribution of dust
can have all possible geometries, although it is expected that the
massive stars in massive clusters will clear the dust from the
cluster's core much more rapidly than they will the peripheries. On
the other hand, the extinction will affect primordially the fainter,
low-mass stars, again adding to the incompleteness effect and probably
contribute, at least in their earlier stages, to aggravate the
incompleteness problem on the cluster scale.

An interesting outcome of the completeness analysis is that in some
cases, such as in the one presented here of a massive cluster with no
interstellar extinction, it may be reasonable to correct the global
luminosity function - and, by extension, the mass function - for
incompleteness, thus gaining information about fainter (low-mass)
regimes that would otherwise be inaccessible, at least using the
completeness tests we experimented with. Other tests may give
different corrections and must be validated beforehand. This opens a
safe, if not new, door to the study of the break of the IMF in massive
galactic and extra-galactic clusters.

 \begin{acknowledgements}
   J. Ascenso acknowledges financial support from FCT, Portugal (grant
   SFRH/BD/13355/2003 and project POCTI/CFE-AST/55691/2004) and is
   grateful to the Calar Alto Observatory for hosting a very important
   part of the work. A special thanks also to Jarle Brinchmann and
   Prof. Pedro Lago for the helpful discussions of statistics. We also
   thank the anonymous referee for the constructive comments that
   contributed to improving this work.
 \end{acknowledgements}

\bibliographystyle{aa}
\bibliography{/Users/Joana/Documents/Work/Docs/bib}

\appendix

\section{Nomenclature} \label{sec:nomenclature}

\parbox{21cm}{The following table summarises the nomenclature used in
  the text.}

\begin{table}[!h]
\begin{minipage}[t]{19cm}
  \caption{}
\label{glossary}
  \begin{tabular}{l p{14.5cm}}
    \hline\hline
    Term & Definition \\
    \hline
    true clusters & Synthetic clusters (Sect. \ref{sec:synthetic-clusters}). \\
    observed* clusters & Synthetic observations (Sect. \ref{sec:synth-observ}). \\
    completeness tests &  Completeness assessment obtained by adding
    artificial stars of increasing magnitude to the cluster image in a
    grid. Derived completeness is defined as the number of detected grid stars with measured
    magnitudes $m_{out}$ that satisfy the condition
    $|m_{out}-m_{in}|<0.1$, divided by the total number of stars of
    magnitude $m_{in}$ in the input grid
    (Sect. \ref{sec:compl-tests}). This would be the assessment an
    observer could do in real data.\\
    true completeness & Completeness assessment obtained by direct
    comparison of the observed* and true clusters. Derived
    completeness is defined as the number of stars in the observed*
    cluster divided by the corresponding number of stars in the true cluster
    (Sect. \ref{sec:global-completeness}). This assessment is only
    possible because we know the true composition of the cluster.\\
    \hline
\end{tabular} 
\end{minipage}
\end{table}

 \end{document}